\documentstyle[preprint,aps]{revtex}
\begin{document}
\tighten
\draft
\title{Stochastic Lattice Gas Model for a Predator-Prey System}
\author{Javier E. Satulovsky\cite{me} 
and T\^ania Tom\'e\cite{tania}}
\address{Instituto de F\'{\i}sica\\
Universidade de S\~ao Paulo\\
Caixa Postal 20516 \\
01452-990 S\~ao Paulo, SP, Brazil}
\date{\today}
\maketitle

\begin{abstract}
We propose a stochastic lattice gas model to describe the dynamics of two
animal species population, one being a predator and the other a prey. This
model comprehends the mechanisms of the 
\mbox{Lotka-Volterra} model. Our analysis
was performed by using a dynamical mean-field approximation and computer
simulations. Our results show that the system exhibits an oscillatory
behavior of the population densities of prey and predators. For the sets of
parameters used in our computer simulations, these oscillations occur at a
local level. Mean-field results predict synchronized collective oscillations.
\end{abstract}

\narrowtext
    
\section{Introduction}
\label{intro}

The dynamics of interacting species has deserved renewed interest since a
simple mathematical model had been
proposed independently by Lotka \cite{lotka} and Volterra \cite{volterra}.
The \mbox{{\it Lotka-Volterra}}
 model predicts an oscillatory temporal evolution of a
predator-prey system and has been widely used as a jumping-off place for
models of many organism societies \cite{goel,nicolis,haken}.
However, some of its appealingly simple assumptions fail 
in taking into account relevant features of such systems.
For instance, 
an individual is generally affected by its local environment 
and not by the global density of each specie. 
The importance of space, 
considered as a limiting resource which would enhance competitive 
coexistence, 
has also been pointed out \cite{sole}. It is in this context 
that lattice gas versions of the model caught recently new attention 
\cite{tainaka1,tainaka2,tainaka3,matsuda} as a suitable way to address these 
questions and simulate practical examples.

We propose a model consisting of a system of two types of interacting
particles residing in the sites of a lattice. One type of particle
represents a prey and the other a predator. Each site can be either empty ($%
0 $), occupied by one prey ($X$) or occupied by one predator ($Y$). The
system evolves in time according to a stochastic irreversible dynamics with
local interactions similar to those occurring in contact
process models \cite{harris,ligget,dickman1} and also in lattice gas models
describing chemical reactions \cite{ziff,aukrust}. 
The local interactions considered
 are: predators can be spontaneously annihilated, 
\mbox{$Y\rightarrow 0$,} 
prey can be autocatalytically created, 
\mbox{$0+nX\rightarrow(n+1)X$,} 
and predators can also be autocatalytically created at the expense of prey, 
\mbox{$X+nY\rightarrow (n+1)Y$.}
In this stochastic treatment of the system, with local evolution
rules, fluctuations inherent to a many-body irreversible interacting system are
taken into account from a microscopic point of view.

We have analyzed the states of the system as a function of the creation
rates of prey and predators and of the death rate of predators considering
square lattices and one dimensional lattices. 
Our computer simulation and mean-field analysis indicate
that the system can exhibit four kinds of states: two absorbing states, one
being a prey-absorbing state and the other a vacuum-absorbing state; and two
active states, one presenting an oscillatory behavior in the population
densities and the other characterized by constant stationary population
densities.

The layout of the paper is as follows. We start in Sec.\ \ref{model} defining 
the model 
and setting up its master equation. In Sec.\ \ref{meanfield} we use mean field  
approximations to solve the system of equations associated to the master 
equation. 
There is a qualitative description of the different type of solutions that we 
have 
found for the one-site and the pair approximations. Sec.\ \ref{simul} 
describes how 
computer simulations were performed and the results that we obtained. A 
summary and 
concluding remarks are stated in 
Sec.\ \ref{final}.

Finally, let us make a short digression on terminology. In this work we use
many times the terms `prey' and `predator'. Eventhough inspired in
mathematical ecology, they are only generic names. Actually, words like
`hosts' or `healthy cells' and `parasites' or `infected cells' may be more 
accurate to mimic a biological process. Or we may simply call them particles 
of type A and B.

\section{The model}
\label{model}

Consider a lattice of $N$ sites which can be either empty, occupied by
a prey or occupied by a predator. At each time step a site is randomly
chosen. For that site, we denote by $n_a$($n_b$) the number of nearest 
neighbors occupied by prey(predators), and by $\zeta$ the total number of 
nearest 
neighbors. Within this model three cases need to be considered:

(a) The site is empty and it becomes either occupied with a prey with a
probability $p_a n_a / {\zeta}$ or else it remains empty with 
probability $1-p_a n_a / {\zeta}$.

(b) The site is occupied by a prey, which is either replaced by a predator
with a probability $p_b n_b / {\zeta}$ or else remains as a prey with 
probability 
$1-p_b n_b / {\zeta}$.

(c) The site is occupied by a predator, which is either vacated with a
probability $p_c$ or else remains as a predator with probability $1-p_c$.

The markovian process defined above involves three parameters: $p_a$, $p_b$
, and $p_c$, which are associated to three subprocesses. Subprocess (a)
describes the birth of prey, subprocess (b) the death of prey and
simultaneous birth of predators, and subprocess (c) the spontaneous death of
predators.

The state of the system is represented by 
\mbox{$\sigma=(\sigma _1,\sigma_2,...,\sigma _N)$} 
where \mbox{$\sigma _i=0,1$} or $2$ 
according to 
whether the site $i$ is empty, occupied by a prey or 
occupied by a predator.

Let $P(\sigma ,t)$ be the probability of state $\sigma $ at time $t$ and let 
$w_i(\sigma )/\tau $ be the probability per unit time of a cyclic
permutation of variable $\sigma _i$. That is, if $\sigma _i=0,1$  or 
$2$, then $w_i$ is the transition probability to $\sigma _i=1,2$ or $%
0$, respectively. The evolution of $P(\sigma,t)$ is governed by the master
equation,
\begin{equation}
\label{master}
\tau \frac d{dt}P(\sigma ,t)=\sum_{i=1}^N\{w_i(\sigma ^i)P(\sigma
^i,t)-w_i(\sigma )P(\sigma ,t)\},
\label{master1} 
\end{equation}
where the state denoted by $\sigma ^i$ is obtained from state $\sigma$ by
an anticyclic permutation of the variable $\sigma _i$.

According to the local rules of the model, defined above, we have:

\begin{mathletters}
\label{trans}
\begin{eqnarray}
w_i(\sigma)&=&p_a\frac
1\zeta \sum_\delta \delta (\sigma _{i+\delta },1)
 \mbox{ if } \sigma_i=0,\\ 
w_i(\sigma)&=&p_b\frac
1\zeta \sum_\delta \delta (\sigma _{i+\delta },2) 
\mbox{ if } \sigma_i=1,\\
w_i(\sigma)&=&p_c\mbox{ if }\sigma_i=2, 
\end{eqnarray}
\end{mathletters}
where the summation is over the $\zeta$ nearest neighbor sites and 
$\delta (x,y)$ is the Kronecker delta.

By rescaling the time the process is found to be invariant under the
transformation $p_a\rightarrow \alpha p_a$, $p_b\rightarrow \alpha p_b$, and 
$p_c\rightarrow \alpha p_c$ where $\alpha$ is a positive constant. Hence we
restrict ourselves to a set of parameters such that they satisfy the
condition $p_a+p_b+p_c=1$ . This is automatically satisfied by writing 
$p_a=1/2-p-c/2$, $p_b=1/2+p-c/2$, and $p_c=c$. All the figures have been made
in terms of parameters $p$ and $c$.

The average of the state function $f(\sigma)$ is defined by:
\begin{equation}
<f(\sigma )>=\sum_\sigma f(\sigma )P(\sigma ,t). \label{aver}
\end{equation}

From Eqs.\ (\ref{trans}) and the master equation\ (\ref{master}), the time
evolution of $<f(\sigma )>$is
\begin{equation}
\tau \frac d{dt}<f(\sigma )>=\sum_{i=1}^N<[f(^i\sigma )-f(\sigma
)]w_i(\sigma )>, 
\label{master2}
\end{equation}
where the state denoted by $^i\sigma $ is obtained from $\sigma $ by a
cyclic permutation of the variable $\sigma _i$.

Let us make the following definitions:
\begin{mathletters}
\label{correl}
\begin{eqnarray}
P_i(\alpha )&=&<\delta (\sigma _i,\alpha )>, \\
P_{ij}(\alpha \beta )&=&<\delta (\sigma _i,\alpha )\delta (\sigma
_j,\beta )>, \\ 
P_{ijk}(\alpha \beta \gamma )&=&<\delta (\sigma _i,\alpha )\delta
(\sigma _j,\beta )\delta (\sigma _k,\gamma )>, \\ 
\cdot&\cdot&\cdot \nonumber
\end{eqnarray}  
\end{mathletters}
where $\alpha ,$ $\beta ,$ and $\gamma $ can take any one of the values $%
0,1,$ or $2$ . Using Eqs.\ (\ref{trans}) we may write the time
evolution equations for some of these probabilities as

\begin{mathletters}
\label{system1}
\begin{eqnarray}
\tau \frac d{dt}P_i(1)&=&p_a\frac 1\zeta \sum_\delta P_{i,i+\delta
}(01) 
-p_b\frac 1\zeta \sum_\delta P_{i,i+\delta }(12), \\
\tau \frac d{dt}P_i(2)&=&p_b\frac 1\zeta \sum_\delta P_{i,i+\delta
}(21)-p_cP_i(2), \\
\tau \frac d{dt}P_{ij}(01)&=&-p_a\frac 1\zeta
\sum_{\stackrel{\delta }{(i+
\delta \neq j)}}(P_{i+\delta
,ij}(101)+P_{ij}(01)) 
+\frac 1{\zeta}\sum_{\stackrel{\delta }{(j+\delta \neq i)}}(
p_aP_{ij,j+\delta }(010)-p_bP_{ij,j+\delta }(012))\nonumber\\
&&+p_cP_{ij}(21), \\
\tau \frac d{dt}P_{ij}(12)&=&\frac 1\zeta 
\sum_{\stackrel{\delta }{
(i+\delta \neq j)}}( p_aP_{i+\delta ,ij}(102)-p_bP_{i+\delta,ij}
(212))+p_b\frac 1\zeta \sum_{\stackrel{\delta }{%
(j+\delta \neq i)}}P_{ij,j+\delta }(112)\nonumber\\
&&-(p_b\frac 1\zeta+p_c)P_{ij}(12), \\
\tau \frac d{dt}P_{ij}(02)&=&p_cP_{ij}(22)-p_a\frac 1\zeta \sum_
{\stackrel{\delta }{(i+\delta \neq j)}}P_{i+\delta ,ij}(102)
+\frac 1\zeta \sum_{\stackrel{\delta }{(j+\delta \neq i)}}p_bP_{ij,j+%
\delta}(012)-p_cP_{ij}(02). 
\end{eqnarray}
\end{mathletters}

These expressions constitute a hierarchic system of equations. The
time evolution of the one-site correlations $P_i(\alpha)$ involve the
two-site correlations $P_{ij}(\alpha\beta)$, the time evolution of the
two-site correlations involve the three-site correlations $P_{ijk}(\alpha
\beta\gamma)$, and so on. A truncation of this hierarchic system, which
we consider in the next section, is the starting point for obtaining
approximate solutions to the problem.

\section{Truncation approximation}
\label{meanfield}

In order to obtain approximate solutions of Eqs.\ (\ref{system1}) we
use a truncation scheme \cite{mamada,kikuchi,dickman2,tome}. 
The simplest truncation scheme is obtained by writing the probability of
a cluster of sites as the product of the probability of each site.
A truncation of higher order, that is of
order $n > 1$, consists in writing any correlation in terms of 
correlations of order $n$ and less than $n$. Consider a cluster
of $m > n$ sites and denote it by $C$. Let $A$ and $B$ be the sets of
points in the `core' and `boundary' of cluster $C$, respectively.
The `core' $A$ is chosen to have $n-1$ sites. The conditional probability
${\cal P}(B|A)$ is approximated by the product 
$\prod_{i \in B}{\cal P}(i|A)$. 
Therefore the probability ${\cal P}(C)$ of cluster $C$ is given by

\begin{eqnarray}
\label{recip}
{\cal P}(C)={\cal P}(A){\cal P}(B|A)&\approx&{\cal P}(A)
\prod_{i \in B}{\cal P}(i|A)={\cal P}(A)\prod_{i \in B}
\frac{{\cal P}(i,A)}{{\cal P}(A)},
\end{eqnarray}
where ${\cal P}(i,A)$ is the probability of the cluster of $n$ sites formed
by site $i$ and the sites of $A$.

\subsection{One-site approximation ($n=1$)}
\label{one}

In this approximation one obtains a closed set of equations for 
the one-site correlation  $P_i(\alpha)$. 
This is accomplished by writing any two-site correlations
$P_{ij}(\alpha\beta)$ as the product $P_i(\alpha )P_j(\beta )$. 
Three one-site correlations are present, 
$P_i(0)$, $P_i(1)$, and $P_i(2)$, 
but only two can be independent. We choose $P_i(1)$ and $P_i(2)$ so that 
$P_i(0)=1-P_i(1)-P_i(2)$. The time evolution of these 
variables are then

\begin{mathletters}
\label{system01}
\begin{eqnarray}
\tau \frac d{dt}P_i(1)&=& p_a\frac 1\zeta P_i(0) \sum_
{\delta }P_{i+\delta }(1)- p_b\frac 1\zeta P_i(1) \sum_
{\delta }P_{i+\delta }(2),\\
\tau \frac d{dt}P_i(2)&=& p_b\frac 1\zeta P_i(2) \sum_
{\delta }P_{i+\delta }(1)- p_c P_i(2).
\end{eqnarray}
\end{mathletters}

Since we seek for homogeneous solutions we write
 $P_i(1)=x$, $P_i(2)=y$ and $P_i(0)=1-x-y$.
Eqs.\ (\ref{system1}) become

\begin{mathletters}
\label{system02}
\begin{eqnarray}
\tau \frac d{dt}x&=& p_a(1-x-y)x - p_bxy,\\
\tau \frac d{dt}y&=& p_bxy - p_cy.
\end{eqnarray}
\end{mathletters}

These equations admit two trivial fixed points, $(x,y)=(1,0)$ and 
$(x,y)=(0,0)$, which correspond to the prey-absorbing and 
vacuum-absorbing states, respectively. The linear stability analysis reveals
 that the latter is always unstable and the former is stable in region I
of the phase diagram (fig.~\ref{phdiagsimple}). Below the critical line, 
defined by \mbox{$c=(2p+1)/3$}, 
 this point becomes unstable giving rise
to a nontrivial stable fixed  point for which $0<x<1$ and 
$0<y<1$, namely an active state. Two regions can be distinguished 
according to the way the nontrivial fixed point is approached. In region II,
the fixed point is an asymptotically stable node (real eigenvalues) whereas 
in region III it is an asymptotically stable focus (complex eigenvalues). 
This implies the emergence of damped oscillations in the population 
densities of the system. The line that separates regions II and III 
is given by $p_ap_c=4p_b(p_b -p_c)$, where $p_a$, $p_b$ and $p_c$ 
are defined in terms of $p$ and $c$ in Sec.\ \ref{model}.

\subsection{Pair approximation ($n=2$)}
\label{two}

This approximation consists in writing the three-site correlations in terms
of two-site and one-site correlations. In this case equation \ 
(\ref{recip})
leads to the following expression for the probability of a cluster of 
three sites,
\begin{equation}
P_{ijk}(\alpha \beta \gamma )=\frac{P_{ij}(\alpha \beta
)P_{jk}(\beta \gamma )}{P_j(\beta )},
\end{equation}
where sites $i$ and $k$ are nearest neighbors of site $j$.

We also seek for spatially homogeneous and isotropic solutions of
Eqs.\ (\ref{system1}). In this
case we may drop the indexes in $P_i(\alpha)$ and $P_{ij}(\alpha\beta)$.
We then have three one-site correlations $P(\alpha)$, $\alpha=0,1,2$ and
nine two-site correlations $P(\alpha\beta)$, $\alpha,\beta=0,1,2$ .
However, only five of them are independent. We choose them to be $P(1)=x$%
, the prey density, $P(2)=y$, the predator density, $P(01)=u$, $%
P(12)=v$, and $P(02)=w$. The equations for these variables are 

\begin{mathletters}
\label{system2}
\begin{eqnarray}
\tau \frac d{dt}x&=&p_au-p_bv, \\
\tau \frac d{dt}y&=&p_bv-p_cy, \\
\tau \frac d{dt}u&=&\frac{(\zeta -1)}\zeta%
 \left(p_a\frac{qu-u^2}z-p_b\frac{uv}x\right)+p_cv-p_a\frac 1\zeta u, \\
\tau \frac d{dt}v&=&\frac{(\zeta -1)}\zeta \left( p_a\frac{uw}z+p_b%
\frac{rv-v^2}x\right)-p_b\frac 1\zeta v-p_cv, \\
\tau \frac d{dt}w&=&\frac{(\zeta -1)}\zeta%
\left(p_b\frac{uv}x-p_a\frac{uw}z\right)+p_c\left( s-w\right),
\end{eqnarray}
\end{mathletters}
where $z=P(0)=1-x-y$, $q=P(00)=z-u-w$, $r=P(11)=x-u-v$, and 
$s=P(22)=y-v-w$.

We have obtained analytically the stationary solutions of the system of
Eqs.\ (\ref{system2}). Due to their somewhat lengthy expressions,
we thought more convenient to make a qualitative discussion here. Their
explicit dependence on the parameters p and c for  a general $\zeta$ is 
given in Appendix A.

For the case $\zeta =2$, the system admits only one stationary solution which is
the prey-absorbing state.

We have analyzed these equations for $\zeta =4$. They have two
trivial fixed points: $(x,y,z,u,v,w)=(1,0,0,0,0)$ and $%
(x,y,z,u,v,w)=(0,0,0,0,0)$ that correspond to the prey-absorbing and
vacuum-absorbing states respectively. The linear stability analysis reveals
that the former becomes unstable in a line of critical points which 
corresponds to the curve 
separating regions I and II in fig.~\ref{phdiagpairs}. 
The region where the prey-absorbing state is stable is denoted by I.
Below this critical line, the system of equations displays a non trivial
fixed point, namely a stationary solution, with $0<x<1$ and $0<y<1$.
Fig.~\ref{pairsprofile} 
shows the densities of prey, predators and empty sites as a
function of c for the case $p=0$.As c decreases
three types of fixed points are obtained, which correspond to regions II,
III, and IV of fig.~\ref{phdiagpairs}. 
Regions II and III differ 
in the way the stable fixed point is approached. Region II has an
asymptotically stable node whereas region III shows an asymptotically
stable focus. At the transition line, fixed points are asymptotically
one-tangent stable focuses. This is due to the emergence of complex
eigenvalues associated to the dominant eigenvector and implies the emergence
of damped oscillations in the population densities of the system.

For sufficiently small values of $c$ a Hopf bifurcation takes place at a
critical line. Inside region IV limit cycles are present as shown in 
fig.~\ref{limitcycle}. These results can also be seen in 
fig.~\ref{bifurcation}, where the real component $\gamma $ 
and the squared imaginary component $\omega ^2$ of the dominant
eigenvalue associated to the focus are plotted against $c$ for $p=0$ . We
also observe the behavior $\omega \sim \sqrt{c}$ for sufficiently small
values of $c$ .

\section{Simulation}
\label{simul}

Numerical simulations were performed in square lattices with periodic
boundary conditions. Unless another size is specified, all the figures in
this section correspond to simulations in lattices of $100\times 100$ sites.

Each run started with an initial configuration of prey, predators and empty
sites placed randomly in the lattice.

Simulation results are summarized in the phase diagram of 
fig.~\ref{phdiagsimul}.

Region I represents prey-absorbing states. Increasing the predator survival
probability (decreasing $c$), the system undergoes a second order kinetic
phase transition towards active stationary states that are represented by
region II in the phase diagram, see fig.~\ref{simulprofile}. 
These states are characterized by short 
transients of less than 1000 MCS. An active state in
the phase diagram means that the lifetime of the system is much longer than
30000 MCS. The states in region II have a constant mean value of the
population densities and our analysis does not indicate any oscillatory
behavior in this region. This behavior can be also seen from the graph of
the spectral density $S(w)\,$of their temporal evolution, shown in 
fig.~\ref{non_osc}. 
Spectral densities were obtained using the Fourier transform (FT) of
temporal samples \cite{vancampen} : 
\begin{equation}
\label{specdens}
S(w_j)=\frac 1N\left\langle \left| \sum_{n=1}^{N}
f(n)\exp \left( iw_jn\right) \right| ^2\right\rangle,  
\end{equation}
\[ w_j=\frac{\pi (j-1)}N, ~~~ j=1,N ~~ ,\]
where $f(n)$ denotes the stationary population density at time $n$. To avoid
a divergence in $S(w)$ at $w=0$ we subtracted to each temporal sample its
mean value, thus obtaining a zero average process. Spectral densities in
 fig.~\ref{non_osc} and fig.~\ref{osc} 
were smoothed by an averaging process over nearest
neighbors.

As $c$ is decreased for a fixed value of $p,$ a transition takes place from
region II to region III, consisting in the emergence of temporal
oscillations in the population densities of the system. The oscillatory
behavior is corroborated by the clear maximum in the spectral density, as
seen in fig.~\ref{osc}.
The degree of oscillation is measured by the eight of the maximum 
value of the spectral density $S(w_{\max })$.
We define a transition point as that in which the low modes of the 
spectral density ($\lim_{w\rightarrow 0}S(w)$) 
become as relevant (have the same value) as its maximum 
$S(w_{\max })$. The transition line between regions II 
and III was determined following this criterion.
Increasing c in region II we get closer to the second order 
kinetic phase transition (KPT). As a consequence, long range
fluctuations become highly relevant and predominance of the low 
modes is to be expected. For this reason, care should be taken 
about the reliability of the
right end of this line (where regions I, II and III
meet in fig.~\ref{phdiagsimul}). Since the line is so close to the
second order KPT, low modes become important before $S(w_{\max })$
suffers a significant decrease, leading to an overlap in $S$.

Simulation results indicate that oscillations become more evident as $c$
decreases, i.e. the maximum of $S(w)\,$ increases as $c\rightarrow 0$. Also $%
w_{\max }$ decreases in this limit, for $p=0\,$ and $c\ll 1$ : $w_{\max
}\sim c^\beta $ with $\beta < 1$. This behavior is in good
qualitative agreement with the pair approximation predictions.

Increasing $c$, $w_{\max }$ also grows, reaching a limit value until the
oscillation disappears. For $p=0$ this value corresponds to a period of
$\approx 250$ MCS.

Region IV in fig.~\ref{phdiagsimul} represents vacuum-absorbing states.
Using lattices up to $480\times 480$ 
the vacuum-absorbing region reduces its height with the
lattice size of the system. This is in agreement with the pair-approximation
predictions that the vacuum-absorbing state occurs only when $c=0$. But it
still remains to be studied the behavior of this region as $L\rightarrow
\infty $, in particular for the case $p > 0$ (where $p_a < p_b$).

In regions I to IV, the system evolves towards an active or absorbing state 
independently of the initial densities of prey and predators. This is not the
case for the parameter set of the grey region in fig.~\ref{phdiagsimul}. In
this region, depending on the initial densities, the system will reach an
oscillating state or will become trapped 
in either of its two absorbing states. 
In the last case, the time it takes for the system to become trapped is much 
less than the transient of any  simulation resulting in an active state. A
possible way to treat both this and the finite size problem (previous
paragraph) is mentioned in Sec.\ \ref{final}.

In order to analyze, from the simulations, whether the system presents
oscillatory synchronized collective states or if the oscillations occur at a
local level, we studied the standard deviation of the temporal samples 
\begin{equation}
\label{22}\sigma =\sqrt{<(\rho -<\rho >)^2>.} 
\end{equation}

For sets of values of the parameters $p$ and $c$ inside
the region III, we have verified that increasing 
the lattice size $L$ of the system $\sigma $ 
decays as $1/\sqrt{N},$ where $N=L^2$ is the number of sites in the
lattice. 
However it still remains to be seen what happens for
values of $c<<1$ . In the hypothetical case 
that $\sigma $ tends to a finite value
as $N\rightarrow \infty $, the system should display synchronized collective
oscillatory states which should correspond to the region IV in 
fig.~\ref{phdiagpairs}, 
as found in the pair approximation analysis .
%Nevertheless, since our model does
%not obey detailed balance, mean field predictions could fail
%\cite{grassberger}.

\section{Conclusions}
\label{final}

We have studied the dynamics of two competing species from a stochastic
point of view developed by using a lattice gas model. A regular square
lattice was considered where each site can be empty, occupied by one prey or
one predator. The system evolves in time according to irreversible
stochastic local rules. The system presents two absorbing states: a
prey-absorbing state and a vacuum-absorbing state. We have found, by means
of a mean field approximation and by computer simulations, that the system
also presents active states where prey and predators are continuously being
created and annihilated. These states can be of two types: a stationary one
and another where populations of the two species oscillate in time. Mean
field results present two kinds of oscillating states: one in which
oscillations are damped and another consisting of a stable limit cycle of
these quantities. The transition from one to the other takes place through a
Hopf bifurcation.

Computer simulations also exhibit a prey-absorbing state, an active state
with stationary populations and another active state which exhibits
oscillations in the population densities. These results are in qualitative
agreement with mean field predictions. In the two regimes where the
probability of spontaneous death of a predator and prey reproduction are
very small, simulations are difficult because the system eventually becomes
trapped into either the prey-absorbing or the vacuum-absorbing state, even
for large systems. This problem could be overcome by using another procedure
in which absorbing states are avoided modifying the model in order to
include a small probability of spontaneous prey creation (of order $1/N$).
Another alternative could be to protect the very
last individual of each specie against extinction, 
avoiding completely the two absorbing states.
We are still testing these procedures.

In the region where simulations are conclusive we have observed that
oscillations occur at a local level revealing that the oscillatory behaviour
displayed by real prey-predator systems might be a local phenomenon.
% this is in agreement with studies in
%other models \cite{grinstein,binder} where it has been argued that no 
%collective states should be expected to exist in the thermodynamic limit.
By this we mean that there is a finite correlation length $\lambda$.
 However we do not discard the possibility of having
synchronized collective oscillations in the system ($\lambda \rightarrow 
\infty$) in the limit where the predator death probability is very small.
 This speculation is based on the fact that mean-field results indicate that 
a limit cycle is  stable in this regime.
 This analysis will be object of a future work.

\acknowledgments
J.E.S. would like to acknowledge the financial support 
 by the FAPESP (Funda\c{c}\~ao de Amparo \`a Pesquisa do Estado de S\~ao Paulo)
 under its fellowship. We also acknowledge FAPESP for support under Projeto 
Tematico \mbox{90/3771-4.}

\appendix
\section{}

As mentioned in Sec.\ \ref{two} we will write the stationary solutions 
of the Eqs.\ (\ref{system2}).

Defining the auxiliary functions:

\begin{eqnarray*}
\alpha&=&\frac\zeta{\zeta-1}\left( \frac cb+\frac{\zeta-2}\zeta\right),\\
\bar{\alpha}&=&\frac\zeta{\zeta-1}\left( \frac{\zeta-2}\zeta-\frac cb\right),\\
\beta&=&1+2\frac ab,\\
\bar{\beta}&=&2\frac ca\frac\zeta{(\zeta-1)},\\
\gamma&=&\frac\zeta{\zeta-1}\left( 1-\frac cb\right),\\
\bar{u}&=&\frac 1{2A}(B-\sqrt{B^2-4AC}),\\
A&=&(\beta-1)(\beta+1),\\
B&=&(\gamma+\bar{\alpha})(\beta+1)+(\beta-1)(\alpha+\bar{\alpha})
   +2\bar{\beta}\beta,\\
C&=&(\gamma+\bar{\alpha})(\alpha+\bar{\alpha})+2\bar{\beta}\bar{\alpha},
\end{eqnarray*}

the solutions are:

\begin{eqnarray}
x&=&(1+\frac ac\bar{u}+\frac{2\bar{u}}{\alpha+%
  \bar{\alpha}-(\beta+1)\bar{u}})^{-1},\\
y&=&\frac acx\bar{u},\\
u&=&x\bar{u},\\
v&=&\frac abx\bar{u},\\
w&=&(\beta\bar{u}-\bar{\alpha})(1-x-y).
\end{eqnarray}

\begin{figure}
\caption{Phase diagram in c-p plane according to
the one-site approximation. Region I represents 
prey-absorbing states. In region II active states are asymptotically stable 
nodes and in region III active states are asymptotically stable focuses.}
\label{phdiagsimple}
\end{figure}

\begin{figure}
\caption{Phase diagram in c-p plane according to the pair approximation. 
Region I represents 
prey-absorbing states. In region II active states are asymptotically stable 
nodes and in region III active states are asymptotically stable focuses. 
Region IV corresponds to limit cycle solutions.}
\label{phdiagpairs}
\end{figure}

\begin{figure}
\caption{Population densities of prey, predators and empty sites according to
 the pair approximation as $c$ varies for $p=0$.}
\label{pairsprofile}
\end{figure}

\begin{figure}
\caption{A typical limit cycle. In this case $p=0$ and $c=0.017$. Curves were 
plotted by iterating the pair approximation eqs. for two different initial 
conditions A and B. Both trajectories are counterclockwise oriented.}
\label{limitcycle}
\end{figure}

\begin{figure}
\caption{Real component $\gamma$ and squared imaginary component $\omega^2$ of 
the dominant eigenvalue associated to the focus of the pair approximation eqs.
, $p=0$ while $c$ varies through regions II, III and IV.}
\label{bifurcation}
\end{figure}

\begin{figure}
\caption{Phase diagram in c-p plane obtained from
simulations for a square lattice of $L=100$. 
Region I represents 
prey-absorbing states, region II non-oscillant active states and region III 
oscillant active states. Region IV consists of vacuum-absorbing states. In 
the grey region no reliable information can be drawn due to the strong initial
configuration dependence.}
\label{phdiagsimul}
\end{figure}

\begin{figure}
\caption{Population densities of prey, predators and empty sites according to 
simulations as $c$ varies for $p=0$.}
\label{simulprofile}
\end{figure}

\begin{figure}
\caption{(a): Temporal evolution of the population densities of prey 
and predators for $p=0$ and $c=0.1795$ (region II). (b): Spectral density of 
the prey population density (similar graphs are obtained for predators and 
empty sites).}
\label{non_osc}
\end{figure}

\begin{figure}
\caption{(a): Temporal evolution of the population densities of prey 
and predators for  $p=0$ and $c=0.022$ (region III). (b): Spectral density
 of the prey population density (similar graphs are obtained for predators and 
empty sites).}
\label{osc}
\end{figure}

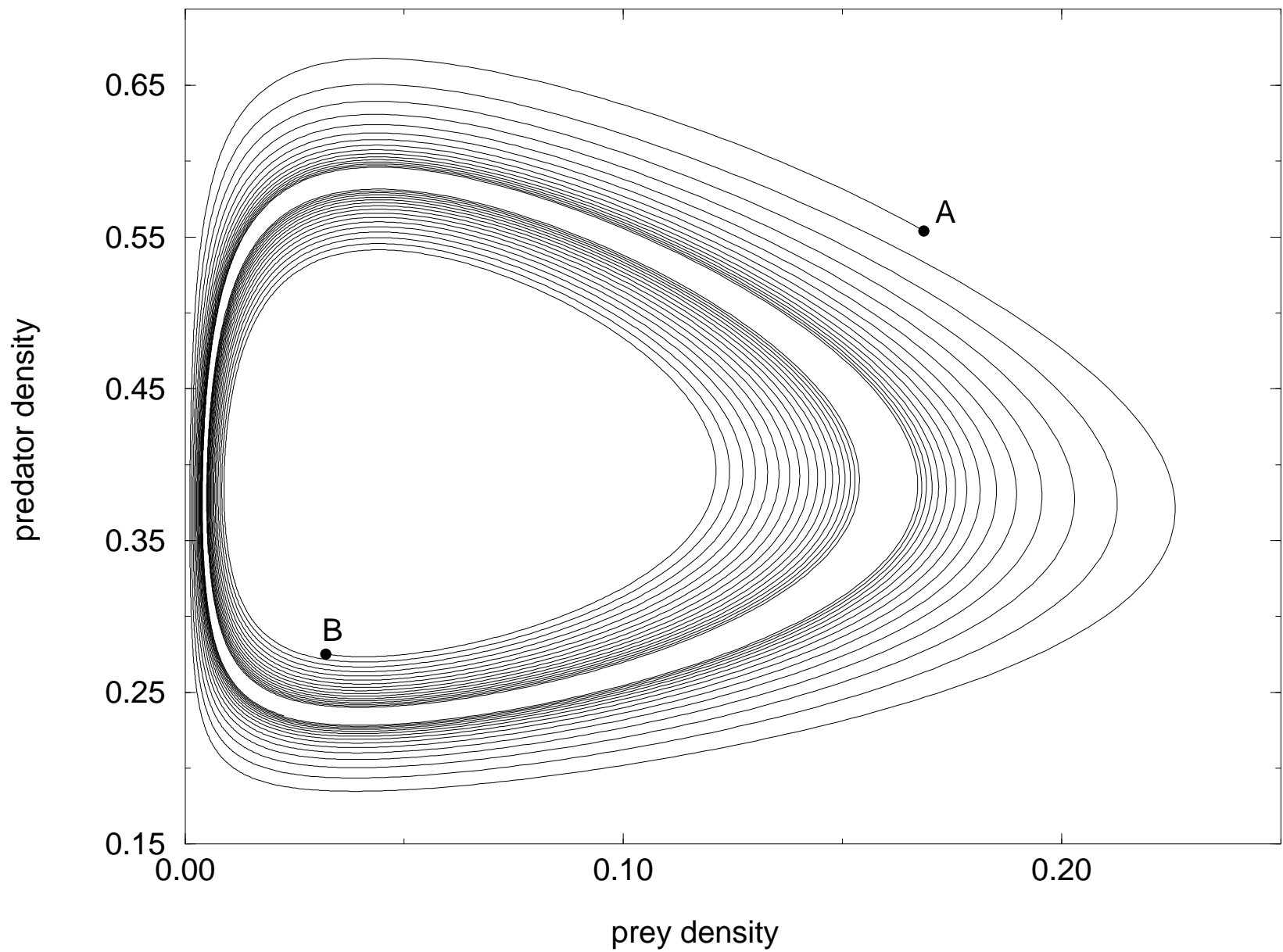
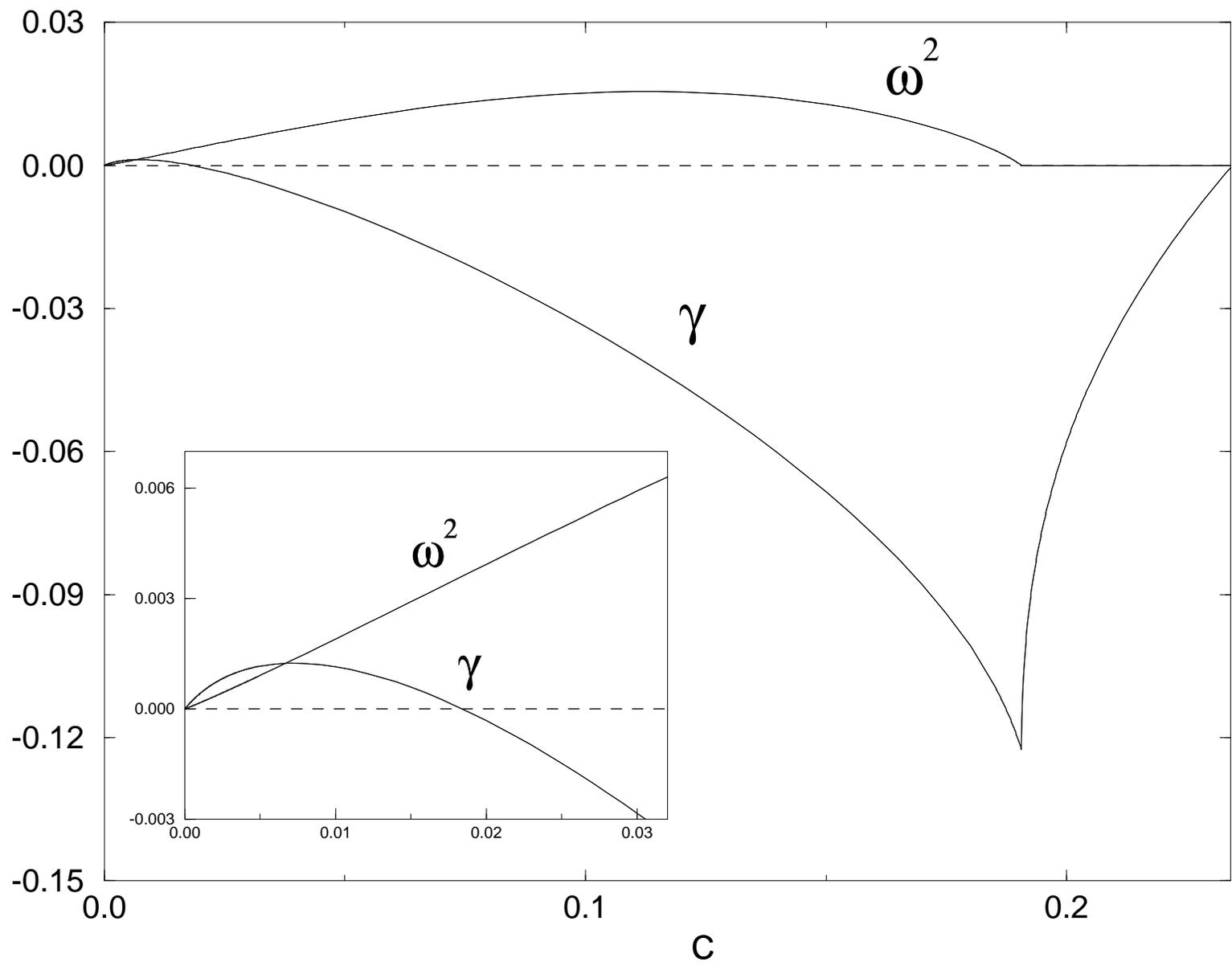
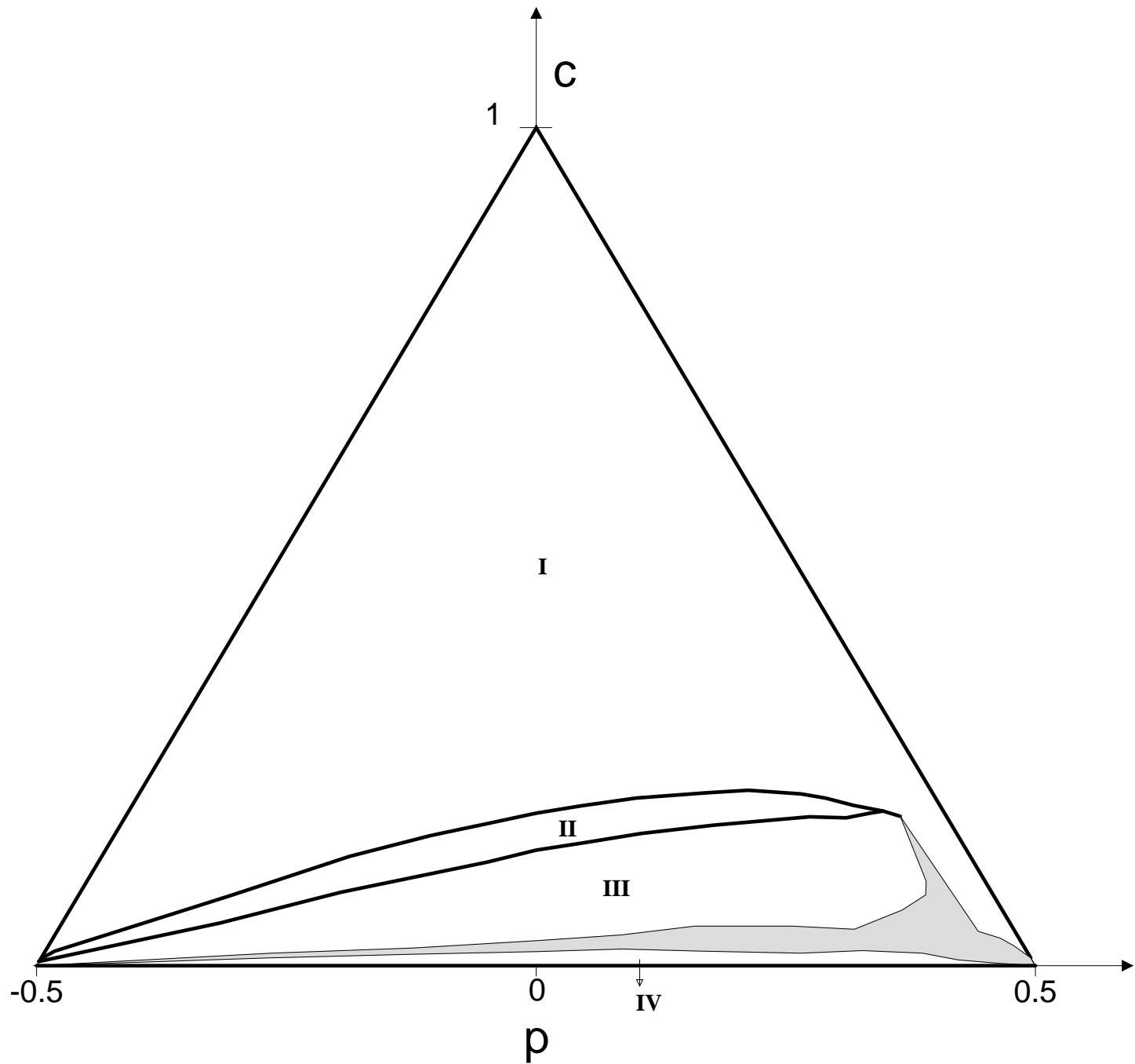
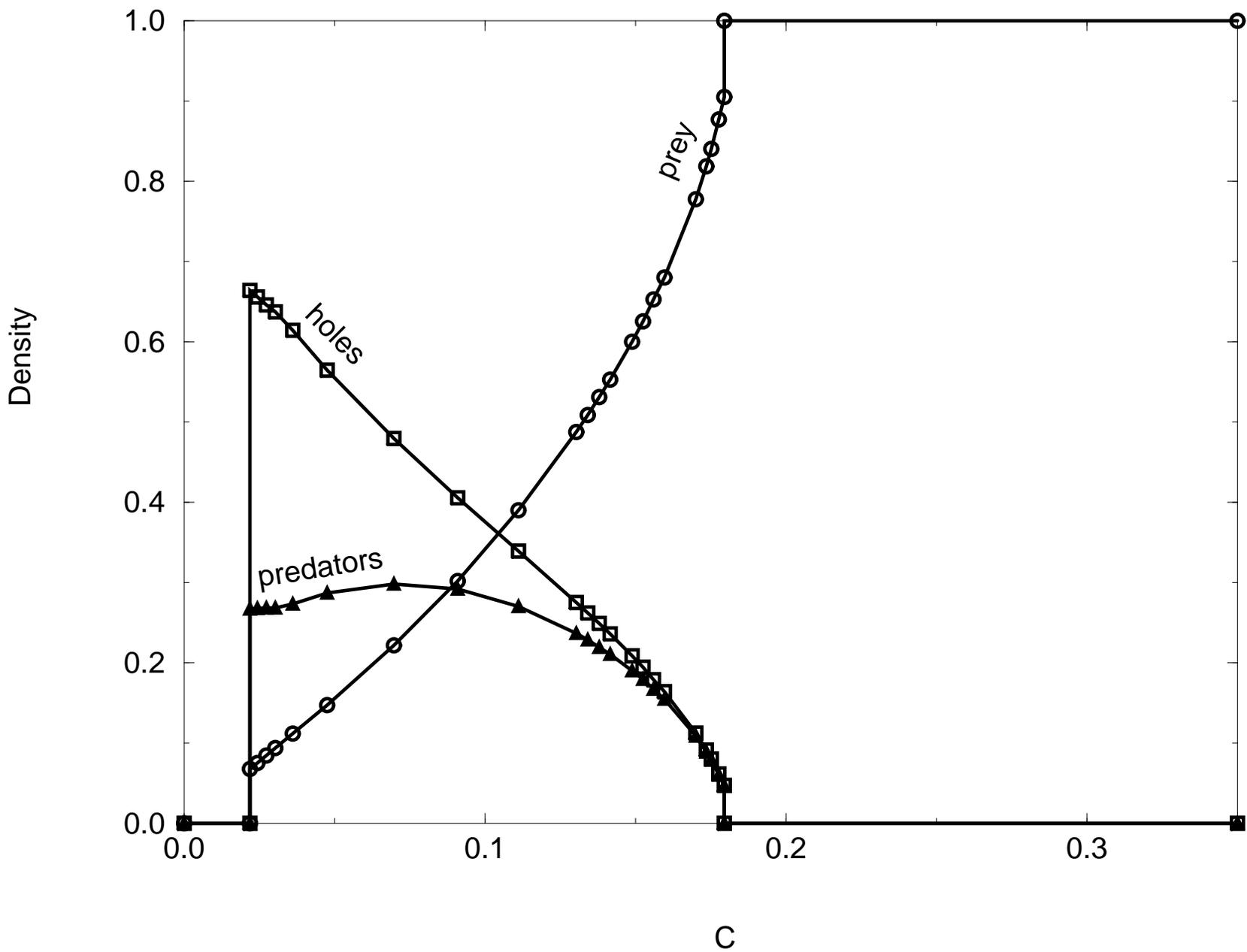
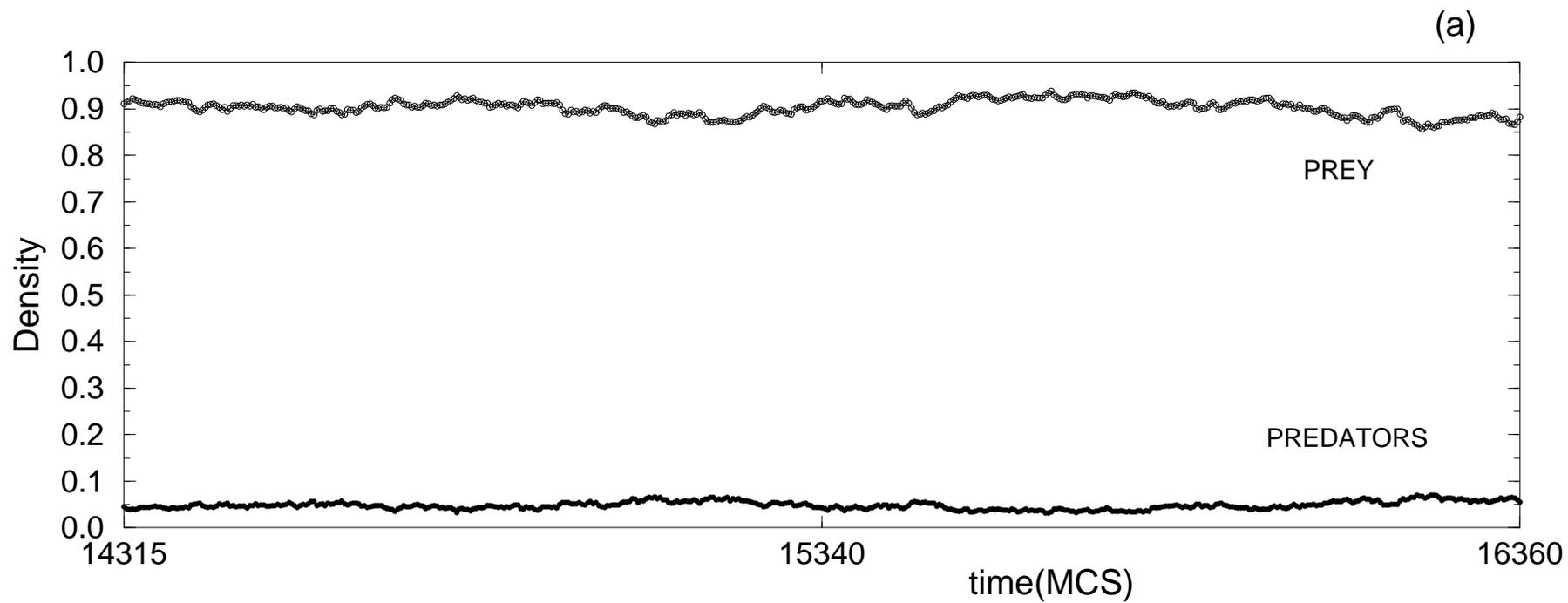
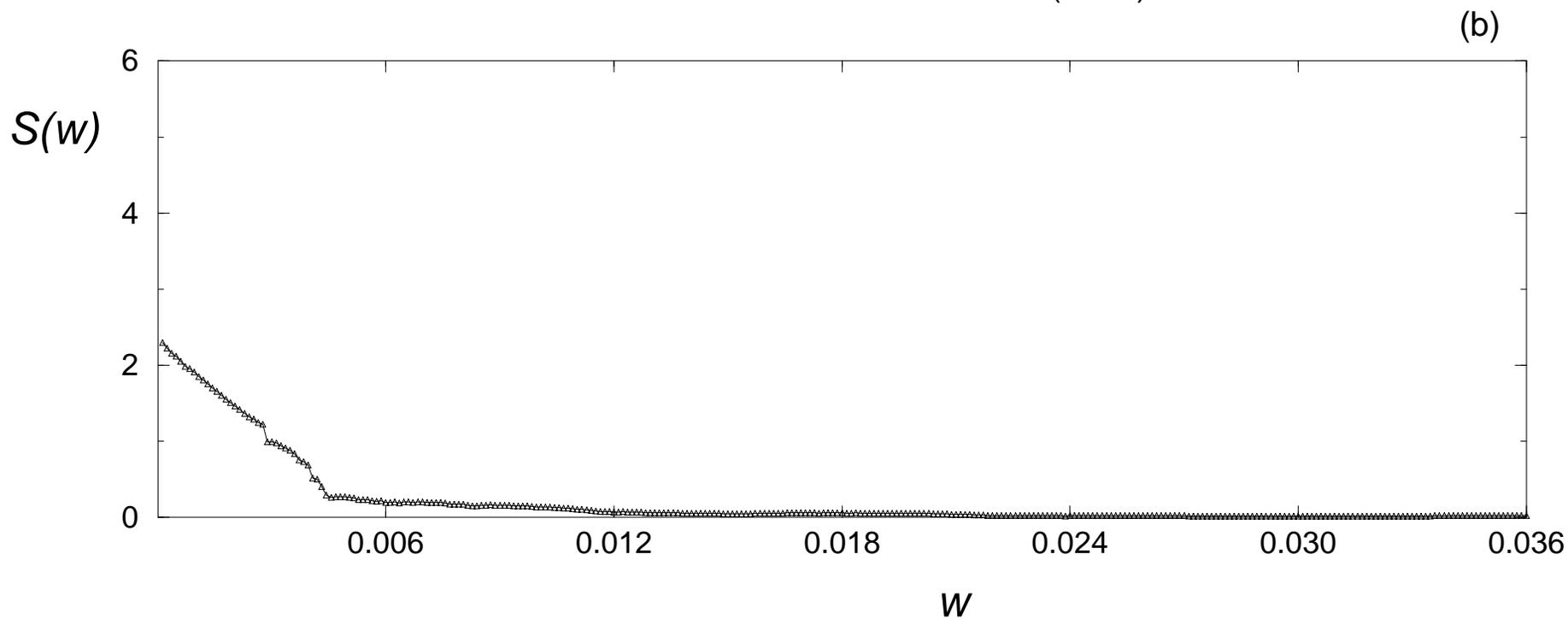
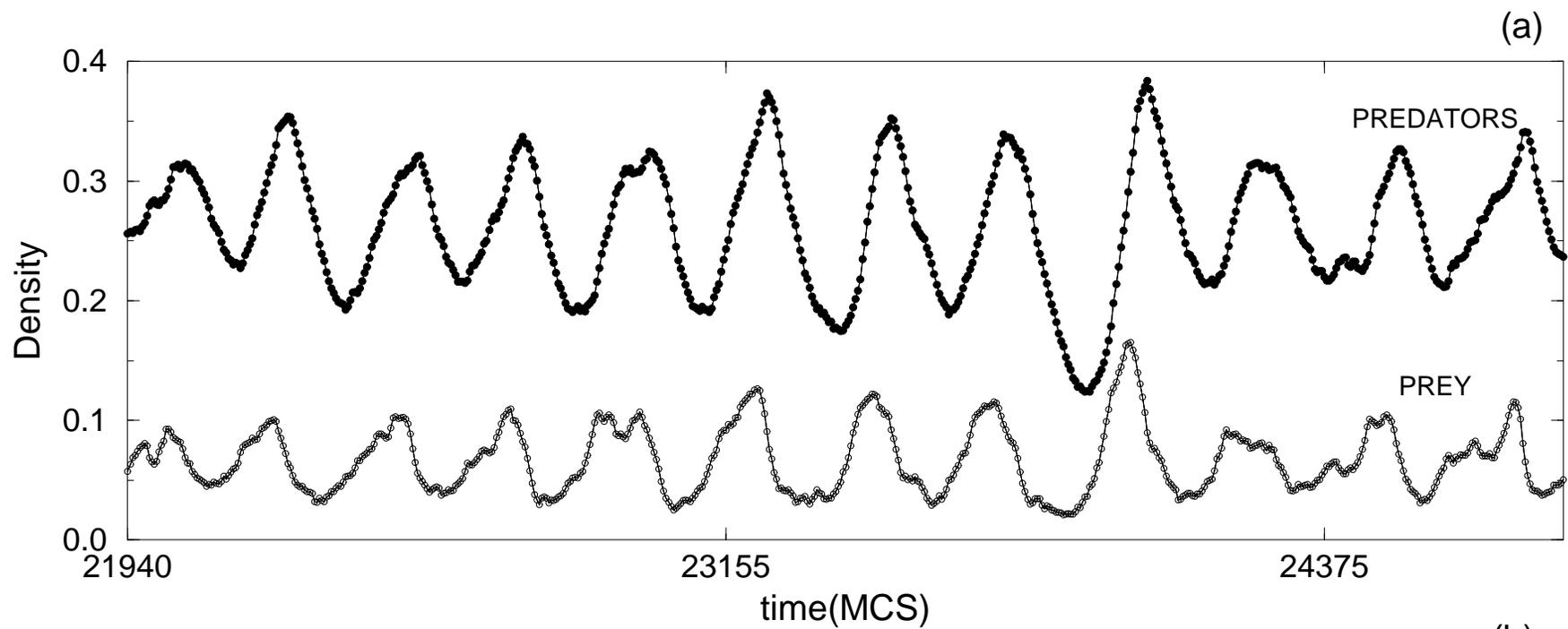
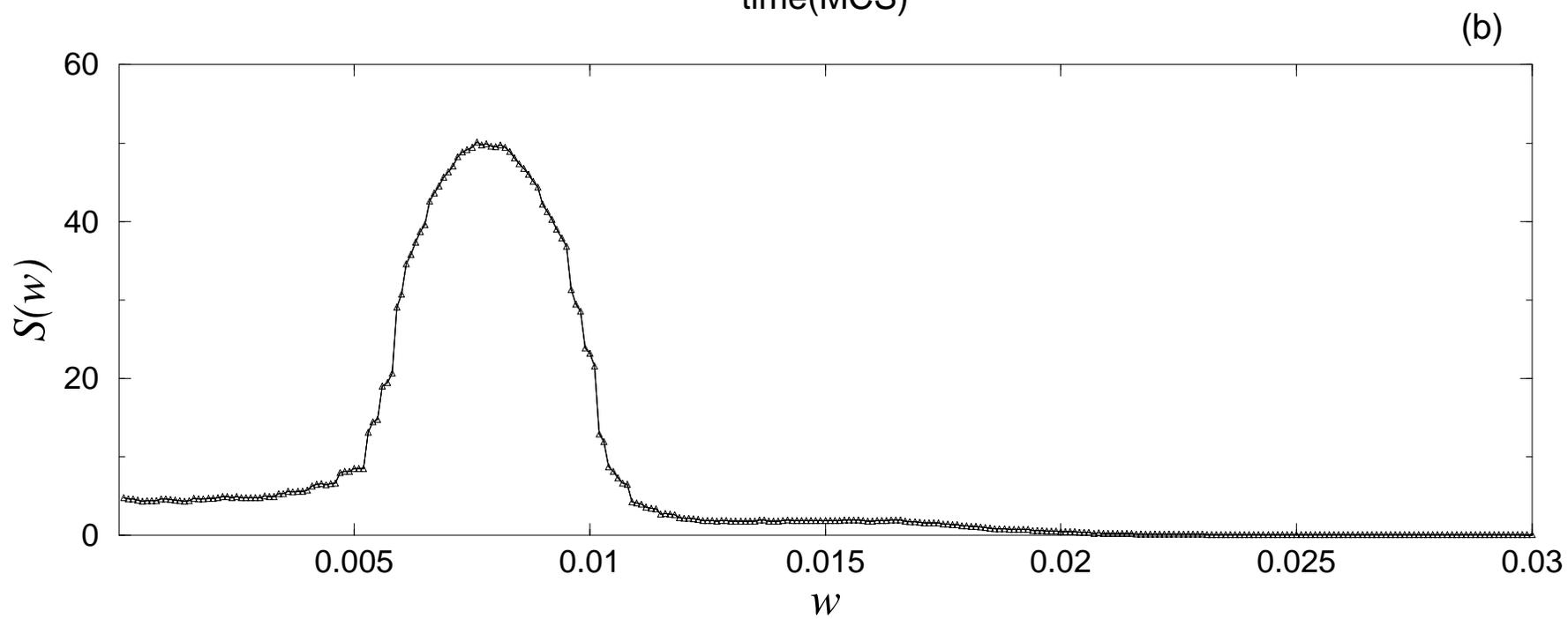
\end{document}